\documentclass[11pt]{article}

\usepackage{a4}
\usepackage[mathscr]{eucal}
\usepackage{citesort,bibmods}
\usepackage{url}
\usepackage{amscd}
\usepackage{amssymb,latexsym}
\usepackage{theorem}      

\usepackage{eucal}                      
\def\greaterthansquiggle{\raise.3ex\hbox{$>$\kern-.75em\lower1ex\hbox{$\sim$}}}
\def\lessthansquiggle{\raise.3ex\hbox{$<$\kern-.75em\lower1ex\hbox{$\sim$}}}

\newcommand{\Cut}{\text{\rm Cut}}
\newcommand{\Span}{\text{\rm Span}}

\newcommand{\ii}{\text{\sl I}\!\text{\sl I}}

\newcommand\fend{{\mbox{\rm \scriptsize end}}}
\newcommand\diff{{\mbox{\rm \scriptsize diff}}}
\newcommand\cHend{\cH_\fend}
\newcommand\cHdiff{\cH_\diff}
\newcommand\hyp{\mycal S}
\newcommand\Gregbeq{\begin{eqnarray}}
\newcommand\Gregeeq{\end{eqnarray}}

\newcommand{\la}{\langle}               
\newcommand{\ra}{\rangle}

\newcommand{\cn}{\colon}        

\newtheorem{thm}{Theorem}
\newtheorem{lemma}[thm]{Lemma}
\newtheorem{prop}[thm]{Proposition}

{\theorembodyfont{\rmfamily} \newtheorem{remark}[thm]{Remark} }

\newcommand{\SAl}{S_{\Al}}

\newcommand{\hs}{\cH_{\mbox{\scriptsize sing}}}

\newcommand{\myremark}{\begin{remark}\rm}
\newcommand{\myendremark}{\end{remark}}

\newcommand{\pr}{\mbox{\rm pr}}

\newcommand{\Hau}{{\mathfrak H}}

\title{On fine differentiability properties of horizons and
applications to Riemannian geometry}

\author{Piotr T.\ Chru\'sciel\thanks{ Supported in part by KBN grant
    \# 2 P03B 130 16. \emph{E--mail}: Chrusciel@Univ-Tours.fr} \\
D\'epartement de
  Math\'ematiques\\ Facult\'e des Sciences\\ Parc de Grandmont\\
  F37200 Tours, France
\\ \\ 
 Joseph H. G. Fu\thanks{Supported in
    part by NSF grant \# DMS-9972094.  \emph{E--mail}:
    fu@math.uga.edu}\\
        Department of Mathematics\\ 
        University of Georgia\\
        Athens, GA 30602, USA
\\  \\Gregory J. Galloway\thanks{Supported in
    part by NSF grant \# DMS-9803566.  \emph{E--mail}:
    galloway@math.miami.edu}\\ Department of Mathematics
\\ University of Miami\\ Coral Gables FL 33124, USA
\\ \\
  Ralph Howard\thanks {Supported in part by DoD Grant \#
    N00014-97-1-0806. \emph{E--mail}: howard@math.sc.edu} \\ Department
  of Mathematics\\ University of South Carolina \\ Columbia S.C.
  29208, USA
}

\newcommand{\U}{{\mathscr  U}}

\newcommand{\cH}{{\mathscr H}}

\newcommand{\Al}{{\mathcal Al}}

\newcommand{\cU}{{\mathscr  U}}
\newcommand{\cV}{{\mathscr  V}}

\newcommand{\cN}{{{\mathscr  N}}{}}

\newcommand{\cC}{{\mathscr  C}}
\newcommand{\cO}{{\mathscr  O}}

\newcommand{\text}[1]{\mbox{\rm #1}}
\newcommand{\qed}{\hfill $\Box$ \medskip}

\newcommand{\kaux}{\sigma}

\newcommand{\Nor}{\text{\rm Nor}}

\newcommand{\e}{\varepsilon}

\newcommand{\beq}{\begin{equation}}

\newcommand{\eeq}{\end{equation}}
\newcommand{\ee}{\end{equation}}
\newcommand{\beqa}{\begin{eqnarray}}
\newcommand{\eeqa}{\end{eqnarray}}
\newcommand{\beqan}{\begin{eqnarray*}}
\newcommand{\eeqan}{\end{eqnarray*}}
\newcommand{\ba}{\begin{array}}
\newcommand{\ea}{\end{array}}

\def\nz{\ifmmode {I\hskip -3pt N} \else {\hbox {$I\hskip -3pt N$}}\fi}
\def\zz{\ifmmode {Z\hskip -4.8pt Z} \else
       {\hbox {$Z\hskip -4.8pt Z$}}\fi}
\def\qz{\ifmmode {Q\hskip -5.0pt\vrule height6.0pt depth 0pt
       \hskip 6pt} \else {\hbox
       {$Q\hskip -5.0pt\vrule height6.0pt depth 0pt\hskip 6pt$}}\fi}
\def\rz{\ifmmode {I\hskip -3pt R} \else {\hbox {$I\hskip -3pt R$}}\fi}
\def\cz{\ifmmode {C\hskip -4.8pt\vrule height5.8pt\hskip 6.3pt} \else
       {\hbox {$C\hskip -4.8pt\vrule height5.8pt\hskip 6.3pt$}}\fi}
\def\au{{\setbox0=\hbox{\lower1.36775ex\hbox{''}\kern-.05em}\dp0=.36775ex\hs
kip0pt\box0}}
\def\ao{{}\kern-.10em\hbox{``}}

{\catcode `\@=11 \global\let\AddToReset=\@addtoreset}
\AddToReset{equation}{section}

\newcounter{mnotecount}[section]

\DeclareFontFamily{OT1}{rsfs}{} 
\DeclareFontShape{OT1}{rsfs}{m}{n}{ <-7> rsfs5 <7-10> rsfs7 <10-> rsfs10}{} 
\DeclareMathAlphabet{\mycal}{OT1}{rsfs}{m}{n} 

\newcommand{\eq}[1]{(\ref{#1})}

\newcommand{\commentout}[1]{}

\newcommand{\cS}{\mathcal{N}^+}

\newcommand{\bea}{\begin{eqnarray}}
\newcommand{\eea}{\end{eqnarray}}
\newcommand{\beaa}{\begin{eqnarray*}}
\newcommand{\eeaa}{\end{eqnarray*}}

\newcommand{\R}{{\mathbb R}}

\begin{document}

\maketitle

\begin{abstract} We study fine differentiability properties of horizons.
  We show that the set of end points of generators of a
  $n$-dimensional horizon $\cH$ (which is included in a
  $(n+1)$-dimensional space-time $M$) has vanishing $n$-dimensional
  Hausdorff measure.  This is proved by showing that the set of end
  points of generators at which the horizon is differentiable has the
  same property. For $1\le k\le n+1$ we show (using deep results of
  Alberti) that the set of points where the convex hull of the set of
  generators leaving the horizon has dimension $k$ is ``almost a $C^2$
  manifold of dimension $n+1-k$'': it can be covered, up to a set of
  vanishing $(n+1-k)$-dimensional Hausdorff measure, by a countable
  number of $C^2$ manifolds. We use our Lorentzian geometry results to
  derive information about the fine differentiability properties of
  the distance function and the structure of cut loci in Riemannian
  geometry.
\end{abstract}

\section{Introduction.}

Horizons are amongst the most important objects that one encounters in
causality theory: Cauchy horizons constitute boundaries beyond which
predictability breaks down; event horizons are boundaries beyond which
no return is possible. The key structural property of horizons is the
existence of {\em generators}: recall that an embedded hypersurface
$\cH\subset M$ is said to be {\em future null geodesically ruled} if
every point $p\in \cH $ belongs to a future inextensible null geodesic
$\Gamma \subset \cH $; those geodesics are called { generators} of
$\cH $. One can then extract the essential properties of Cauchy
horizons, or black hole event horizons, in the following definition:
$\cH$ is \emph{a future horizon} if $\cH $ is an \emph{achronal,
  closed, future null geodesically ruled topological hypersurface}. It
follows from the above definition (or from the properties of past
Cauchy horizons, or from the properties of future event horizons) that
the generators can have past endpoints on $\cH $, but no future
endpoints.

The set $\cHend$ of end points of generators of $\cH$ provides an
important tool in the study of the structure of horizons; for
simplicity we will refer to those points as {\em end points}.  In
particular one wants to know how ``large'' this set can be. One
defines the multiplicity $N(p)$ of a point $p\in\cH$ as the number of
generators which pass through or exit $\cH$ at $p$; it is well known
that if $N(p)>1$, then $p$ is necessarily the endpoint of all relevant
generators. The set of points with multiplicity $N(p)>1$ determines
the differentiability properties of horizons: as has been shown by
Beem and Kr\'olak \cite{BK2}, horizons are non-differentiable
precisely at this set. It is also well known that the set of points at
which a horizon is non-differentiable has vanishing $n$-dimensional
Hausdorff measure, and this gives one control over the size of the set
of endpoints with multiplicity $N(p)>1$.  Thus, in order to control
the dimension of $\cHend$ it remains to estimate that of the set of
endpoints with multiplicity $N(p)=1$.  Let us denote by $\cHdiff$ the
set of points of $\cH$ at which $\cH$ is differentiable; what has been
said shows that the set of endpoints with multiplicity $N(p)=1$
coincides with the set $\cHend\cap\cHdiff$.  Kr\'olak and Beem
\cite{BK2} have displayed an example of a horizon with an end point
with multiplicity one (at which $\cH$ is, of course, differentiable),
thus there exist horizons for which the set $\cHend\cap\cHdiff$ is not
empty.  Our first main result is the following:

\begin{thm}\label{th:a} Let $\cH$ be a future horizon in an
  $(n+1)$-dimensional spacetime $(M,g)$.  Then the set $\cHend \cap
  \cHdiff$ has vanishing $n$-dimensional Hausdorff measure.
  Moreover, for any $C^2$ spacelike hypersurface $\hyp$ the set
  $\cHend \cap \cHdiff \cap \hyp$ has vanishing
  $(n-1)$-dimensional Hausdorff measure.
\end{thm}

The set of points where the multiplicity $N(p)$ is large has a more
precise structure.  To describe this, equip $M$ with an auxiliary
complete Riemannian metric $\kaux$, and for each $p\in \cH$ let
$\cS_p$ be the set of future pointing $\kaux$-unit vectors that are
tangent to a generator of $\cH$ at $p$. We call such vectors
\emph{semi-tangents} to $\cH$. Then the number of vectors in $\cS_p$
is just the multiplicity $N(p)$ of $p$.  Define
$$
\cC_p:=\text{convex cone generated by $\cS_p$}\ .
$$
We can measure the size of the set of generators through $p$ by
$\dim(\cC_p)$ (which is the dimension of the linear span of $\cC_p$ in
$T_pM$).  This is a different measure than is $N(p)$; in particular
this gives finer information when $N(p)=\infty$.  We also set
\begin{equation}
  \label{inclu}
\cH[k]:=\{p : \dim(\cC_p)\ge k\}.
\end{equation}
For $k=1$, $\cH[k]=\cH$ as every point is on at least one generator.
For $k=2$, $\cH[k]$ is the set of points of $\cH$ that are on more
than one generator.  As $\dim (\cC_p)$ is the dimension of the span of
$\cS_p$ and $\cS_p$ contains $N(p)$ vectors, $\dim(\cC_p)\le N(p)$.
This implies
$$
\cH[k]\subseteq \{p\in \cH : N(p)\ge k\}\ .
$$
Also, for $1\le k\le 3$, any $k$ distinct elements of $\cS_p$ are
linearly independent.\footnote{The linear span of two future pointing
  $\kaux$-unit null vectors is a two dimensional timelike subspace and
  there are only two future pointing null rays in this subspace.  So
  a third future pointing $\kaux$-unit vector can not be in the span
  of the first two.}  Therefore if $1\le k\le 3$ and $N(p)\ge k$ then
choosing $k$ distinct, and thus linearly independent, elements of
$\cS_p$, shows that $\dim(\cC_p)\ge k$. Whence
$$
\cH[k]=\{p\in \cH : N(p)\ge k\}\qquad \text{for $1\le k \le 3$}\ .
$$
Our next main result, based on the deep
results\footnote{\label{Albwarn}The reader is warned that the
  dimension index $k$, in $\cH[k]$, is shifted by one, as compared to
  that used in Theorem~3 of \cite{Alberti:sing}; compare the remarks
  following Definitions~1.5 and 1.7 in \cite{Alberti:sing}.} in
\cite{Alberti:sing}, is that the sets $\cH[k]$ are ``almost $C^2$
submanifolds of dimension $n+1-k$, up to singular sets of lower
dimension''.\footnote{We note a related result of~\cite{HusaWinicour},
  where it is shown that the set $\{N(p)=2\}$ is, up to a lower
  dimensional set, a smooth submanifold of co-dimension two for the
  horizons considered there.}  To make this statement precise, let
$\Hau^m$ be the $m$-dimensional Hausdorff measure on $M$ (defined with
respect to some Riemannian metric $\kaux$ on $M$).
Recall,~\cite[Def.~1.1 p.19]{Alberti:sing} a Borel set $\Sigma\subset
M$ is a \emph{$(\Hau^m,m)$ rectifiable set of class $C^2$} iff
$\Sigma$ can be covered, up to a set of vanishing $\Hau^m$~measure, by
a countable collection of $m$-dimensional $C^2$ submanifolds of $M$.
This definition is independent of the choice of the Riemannian metric
$\kaux$.  Following~\cite{Alberti:sing} we will shorten ``$(\Hau^m,m)$
rectifiable set of class $C^2$'' to ``$C^2$~rectifiable of dimension
$m$''.

\begin{thm}\label{th:sing}  For
  $1\le k\le n+1$ the set $\cH[k]$ is a $C^2$~rectifiable set of
  dimension~$n+1-k$.  Therefore $\cH[k]$ has Hausdorff dimension $\le
  n+1-k$.
\end{thm}

Using $k=1$, and that $\cH[1]=\cH$, this implies that horizons are
$C^2$ rectifiable of dimension~$n$.  As they are also locally
Lipschitz graphs they have the further property that $\Hau^n(\cH\cap
K)<\infty$ for all compact sets $K\subseteq M$. When $k=n+1$ this
implies that $\cH[n+1]$ is a countable set (\emph{cf.\/} Remark~\ref{top-bot}
below).

\section{Proof of Theorem~\ref{th:a}}
We shall prove the second part of the theorem; the first part follows
immediately from the second and the co-area formula.  Let, then,
$\hyp$ be as in Theorem \ref{th:a}; since the result is purely local,
without loss of generality we may assume that $\hyp$ is the level set
$\{t=1\}$ of a time function $t$, with range $\mathbb R$, the level
sets of which are Cauchy surfaces.  We use the constructions and
notations of \cite{Ch+}, with $\Sigma_1 = \hyp$ and $\Sigma_2=
\{t=2\}$.  Let
$$
\hat S_1 = \hyp \cap \cHend \cap \cHdiff \, ,
$$
and let $A$, $\phi$ be defined as at the beginning of the proof of
Theorem 6.1 in \cite{Ch+}.  Hence, $A$ is the subset of $S_2 =
\Sigma_2 \cap \cH$ consisting of those points in $S_2$ that are met by
the generators of $\cH$ that meet $S_1 = \Sigma_1 \cap \cH$, and
$\phi: A\to S_1$ is the map that moves the points of $A$ back along
these generators to $S_1$.  We can choose the auxiliary Riemannian
metric $\sigma$ on $M$ so that $dt$ has unit length with respect to
this metric.  Then $A$ is an $A_{\delta}$ set as defined by
Equation~(6.6) of~\cite{Ch+}, with $\delta = 1$.  We set
$$
\hat A = \phi^{-1}(\hat S_1)\subset A \subset S_2 \, ;
$$
thus, the points in $\hat A$ are precisely those points on $\Sigma_2\cap
\cH$ the generators
through which exit $\cH$, when followed to the past, at the differentiable
end points on $\hyp$.

For $i=1,2$, let $\Hau^{n-1}_{h_i}$ denote the $(n-1)$-dimensional
Hausdorff measure on $\Sigma_i$ with respect to the distance function
determined by the induced metric $h_i$ on $\Sigma_i$.  By a
straightforward extension of the proof of \cite[Proposition~6.14]{Ch+}
one has for any $\Hau^{n-1}_{h_2}$-measurable subset $\Omega$ of $A$,
\beq\label{eq:a} \int_{S_1} N(p,S_2)\mathbf{1}_{\phi(\Omega)} d
\Hau^{n-1}_{h_1}(p)= \int_{\Omega} J(\phi)(q)\,
d\Hau^{n-1}_{h_2}(q)\;, \eeq where $\mathbf{1}_U$ denotes the
characteristic function of the set $U$, and $J(\phi)$ is, in a
suitably defined sense (\emph{cf.\/} \cite{Ch+}, Prop.  6.14), the Jacobian of
the locally Lipschitz function $\phi$.

In Proposition \ref{th:b} below we show that there exists a
$\Hau^{n-1}_{h_2}$-negligible set $\hat A'\subset \hat A$ such that
$J(\phi)(q) = 0$ for all $q\in \hat A \setminus \hat A'$.  It then
follows that, \beq \hat A \subset \Omega \equiv\{q\in A : J(\phi)= 0\}
\cup \hat A' \, .  \eeq $\Omega$, as defined above, is the union of a
$\Hau^{n-1}_{h_2}$-measurable set  and a $\Hau^{n-1}_{h_2}$-negligible
set, and hence is itself $\Hau^{n-1}_{h_2}$-measurable.  Equation
\eq{eq:a} then shows that $\phi(\Omega)$ is
$\Hau^{n-1}_{h_1}$-negligible.  Now, since $\hat S_1 \subset
\phi(\Omega)$, the result follows.  \qed

It thus remains to establish the following.

\begin{prop}\label{th:b} $J(\phi)(q) = 0$ for $\Hau^{n-1}_{h_2}$-almost all
$q\in \hat A$.
\end{prop}

\begin{proof}  We use the definitions, constructions and notations of the
proof
of \cite[Prop. 6.16]{Ch+}.   Thus, let $\cU\subset \Sigma_2$ be a coordinate
neighborhood of the form $\cV\times (a,b)$, with $\cV\subset \R^{n-1}$
and $a,b,\in \R$, in which $\cU\cap N$ is the graph of a
$C^{1,1}$ function $g:\cV\to\R$, and in which $\cH\cap \cU$ is the
graph of a semi--convex function $f:\cV\to\R$.  Here $N = N_{\delta}$
is a locally $C^{1,1}$ hypersurface in $\Sigma_2$ into which $A$
has been embedded.  Let $\pr A$ denote the projection onto $\cV$ of
$A\cap\cU$, thus $A\cap\cU$ is the graph of $g$ over $\pr A$.

Now, let $x_0\in \hat B \cap \pr \hat A$, where $\hat B$ is the set of
full measure in $\pr A$ constructed in the proof of \cite[Prop.
6.16]{Ch+}, and $\pr \hat A$ is the projection onto $\cV$ of $\hat
A\cap\U$.  Since $g$ is Lipschitz, the graph of g over $\hat B\cap \pr
\hat A$ has full measure in $\hat A$.  Let $q_0 = (x_0,f(x_0))\in \hat
A$ be the corresponding point on $\cH\cap \Sigma_2$, thus the
generator $\Gamma$ of $\cH$ passing through $q_0$ exits the horizon at
a point $p \in \hyp \cap \cHdiff$.  Let $\hat \Gamma$ be any null
geodesic which extends $\Gamma$ to the past, and let $p_n$ be any
sequence of points on $\hat \Gamma$ which are to the causal past of
$p$ and which approach $p$ as $n$ tends to infinity.  Since $\hat
\Gamma$ is a null geodesic which exits $\cH$ at $p$, the $p_n$'s lie
to the timelike past $I^-(\cH)$ of $\cH$.  Thus the integral curve
$\gamma_n$ of $\frac{\partial}{\partial t}$ starting at $p_n$ meets
$\cH$ at some point $r_n\in \gamma_n \cap \cH$.  One can then
construct a causal curve from $p_n$ to $\cH \cap \Sigma_2$ by
following $\gamma_n$ from $p_n$ to $r_n$, and any generator of $\cH$
passing through $r_n$; as $(M,g)$ is globally hyperbolic this
generator will necessarily intersect $\Sigma_2$.  It follows that
there exists a timelike curve $\hat \gamma_n \subset J^-(\cH)$ from
$p_n$ to $\cH \cap \Sigma_2$.  By the compactness of the space of
causal curves, passing to a subsequence if necessary, the $\hat
\gamma_n$'s converge (in a well known sense) to a causal curve
$\gamma$ from $p$ to a point $q\in \cH \cap \Sigma_2$.  The
achronality of $\cH$ shows that $\gamma$ is a generator of $\cH$
passing through $p$, hence $\gamma = \Gamma$ and $q = q_0$.

Now suppose that $J(\phi)(q_0)\ne 0$.  By the construction of the set
$\hat B$, there exists $g_j\in C^2(\cV)$, approximating $g$, such that
$S_j$, the graph of $g_j$, is a $C^2$ hypersurface in $\Sigma_2$
which, in a well defined sense, makes second order contact with $\cH
\cap \Sigma_2$ at $q_0$.  (More precisely, $g_j$ and $f$, as well as
their first derivatives, agree at $q_0$, and the second derivative of
$g_j$ agrees with the so-called second Alexandrov derivative of $f$ at
$q_0$.)  Since $S_j$ is tangent to $\cH \cap \Sigma_2$ at $q_0$, the
null geodesic $\hat \Gamma$ is normal to $S_j$ at $q_0$.  Let $\phi_j
: S_j\to \hyp$ be the $C^1$ map which moves the points of $S_j$
along the family of null geodesics normal to $S_j$ which includes
$\hat \Gamma$.  Then we have $J(\phi_j(q_0))= J(\phi)(q_0) \ne 0$,
\emph{cf.\/} \cite[Equation~(6.40)]{Ch+}.  Define $g_{j,\e}\in C^2(\cV)$ by
\beq g_{j,\e}(x) = g_j(x) + \e |x-x_0|^2 \, , \eeq and let $S_{j,\e}$
be the graph of $g_{j,\e}$; for $\e >0$, $S_{j,\e}\subset J^+(\cH)$
and $S_{j,\e}\setminus \{q_0\} \subset I^+(\cH)$.  Note that as
$S_{j,\e}$ is tangent to $S_j$ at $q_0$, $\hat \Gamma$ is normal to
$S_{j,\e}$ at $q_0$.

The fact that the Jacobian of $\phi_j$ is nonzero at $q_0$ implies
that $p$ is not a focal point to $S_j$ along $\hat \Gamma$. Moreover,
there can be no focal points to $S_j$ along the segment of $\hat
\Gamma$ to the future of $p$, {\em cf.\/}~\cite[Lemma~4.15]{Ch+}. It
follows that by taking $\e$ small enough and $\hat \Gamma$ short
enough, there will be no focal points to $S_{j,\e}$ along $\hat
\Gamma$.  This implies by normal exponentiation that there exists an
embedded $C^2$ null hypersurface $\cN_{j,\e}$ which contains $\hat
\Gamma$ and, by shrinking it if necessary, $S_{j,\e}$, as well, {\em
  cf.\/} \cite[Prop.  A3]{Ch+}.  Moreover, there exists a neighborhood
$\cO$ of $\hat \Gamma$ in which $\cN_{j,\e}$ is achronal: Indeed,
since spacetime is time orientable, $\cN_{j,\e}$ is a two-sided
connected embedded hypersurface in $M$.  As such $\cN_{j,\e}$ admits a
connected neighborhood $\cal O$ which is separated by $\cN_{j,\e}$
($\cN_{j,\e} \subset \cal O$, and ${\cal O }\setminus \cN_{j,\e}$
consists of two components).  Then a future directed timelike curve
joining points of $\cN_{j,\e}$ would be a timelike curve from the
future side of $\cN_{j,\e}$ to the past side of $\cN_{j,\e}$, which is
impossible if the curve remains in $\cal O$. We conclude that
$\cN_{j,\e}$ is achronal in $\cal O$.

Consider now the timelike curves $\hat\gamma_n \subset J^-(\cH)$
constructed earlier in the proof; since the $\hat \gamma_n$'s converge
to $\Gamma$ there exists $n_0$ such that all the $\hat\gamma_n$'s are
entirely contained in $\cO$ for $n\ge n_0$.  Moreover, by taking $n_0$
larger if necessary, it is clear that each such $\hat\gamma_n$ will
meet the hypersurface $P$ in $M$ obtained by pushing $S_{j,\e}$ to the
past along the integral curves of $-\frac{\partial}{\partial t}$.  One
can then construct a timelike curve from $p_n$ to $S_{j,\e}$ contained
in $\cO$ by following $\hat \gamma_n$ from $p_n$ to $P$, and then an
integral curve of $\frac{\partial}{\partial t}$ to $S_{j,\e}$.  This
contradicts the achronality of $\cN_{j,\e}$ in $\cO$, and establishes
Proposition \ref{th:b}. \qed
\end{proof}

\section{Proof of Theorem~\ref{th:sing}}

We start by showing that $\cH$ has no worse regularity than being the
boundary of a convex set:

\begin{prop}\label{convex-graph}
  For any point $p\in\cH$ there is a coordinate system $x^1,\ldots,
  x^{n+1}$ defined on an open set $U\subseteq M$ so that $\cH\cap U$
  is given by the graph $x^{n+1}=h(x^1,\ldots, x^n)$ of a convex
  function $h$.
\end{prop}

\begin{proof}
  It is shown in~\cite[Theorem~2.2]{Ch+} that $\cH$ is locally the
  graph of a semi-convex function.  That is, there is a coordinate
  system $y^1,\ldots, y^{n+1}$ so that $U\cap \cH$ is given by a graph
  $y^{n+1}=u(y^1,\cdots,y^n)+h(x^1,\cdots,x^n)$ where $u$ is
  $C^\infty$ and $h$ is convex.  Define new coordinates by $x^i=y^i$
  for $i=1,\ldots,n$ and $x^{n+1}=y^{n+1}-u(x^1,\ldots,x^n)$.  In
  these coordinates $\cH$ is given by $x^{n+1}=h(x^1,\ldots,
  x^n)$.\qed
\end{proof}

Recall that a convex body in a finite dimensional vector space has a
well defined normal cone at each of its boundary points $p$.  One
definition of this tangent cone is the set of linear functionals on
the vector space such that their restrictions to the body attain a
maximum at the point $p$.  The following is an adaptation of this
definition to manifolds which allows us to define the tangent cones
$\Nor_p(J^+(\cH))$ to $J^+(\cH)$ at points $p\in \cH=\partial
J^+(\cH)$ in an invariant manner.
\begin{eqnarray*}
  \Nor_p(J^+(\cH))&:=&\{df(p) : \text{$f\in C^\infty( M,\R)$  and}
\\ & & \text{$f\big|_{J^+(\cH)}$ has a local max.\ at $p$}\}\ .
\end{eqnarray*}

The following is a special case of a main
result$^{\mbox{\rm\scriptsize \ref{Albwarn}}}$ in the paper of
Alberti~\cite[Thm~3 p.~18]{Alberti:sing} adapted to our notation.

\begin{prop}[Alberti]\label{AO}
Let $B$ be convex body in an $n+1$-dimensional vector space.  For each
$k=1,2,\ldots,n+1$ let $\partial B[k]$ be the set of points $p\in
\partial B$ so that $\dim\Nor_p(B)\ge k$.  Then $\partial B[k]$ is a
$C^2$ rectifiable set of dimension~$n+1-k$.  Therefore the Hausdorff
dimension of $\partial B[k]$ is less then or equal to $n+1-k$.\hfill $\Box$
\end{prop}

\begin{remark}\label{top-bot}
  The top and bottom dimensional cases of this are worth remarking on.
  When $k=n+1$ this implies that $\partial B[n+1]$ is a $C^2$
  rectifiable set of dimension~$0$.  But then, \cite[Theorem~3 and
  Definition~1.1]{Alberti:sing}, $\partial B[n+1]$ is a countable
  union of sets of finite $\Hau^0$ measure.  However the zero
  dimensional measure $\Hau^0$ is just the counting measure,
  \cite[p.~171]{Federer:book}, so that $\Hau^0(A)$ is just the number
  of points in $A$. Therefore $\partial B[n+1]$ is countable as it is
  a countable union of finite sets.  Consider, next, $k=1$; as
  $\dim\Nor_p(J^+(\cH)) \ge 1$ for all $p\in \partial B$ we have
  $\partial B=\partial B[1]$ and therefore $\partial B$ is a $C^2$
  rectifiable set of dimension~$n$.  Because $\partial B$ is also
  locally the graph of a Lipschitz function it has the further
  property that $\Hau^n(K\cap \partial B)<\infty$ for all compact sets
  $K$.  The corresponding fact is not true for $\partial B[k]$ when
  $2\le k\le n-1$ (cf.~\cite[Thm.~2 p.~18]{Alberti:sing}).
\end{remark}

Theorem~\ref{th:sing} follows immediately from this,
Proposition~\ref{convex-graph}, and

\begin{lemma}\label{cone}
For each $p\in \cH$ the normal cone $\Nor_p(J^+(\cH))$ is given by
$$
\Nor_p(J^+(\cH))=\{\langle v,\cdot\rangle : v\in \cC_p\}\ .
$$
\end{lemma}

\begin{proof} 
We choose a coordinate system $x^1,\ldots,x^{n+1}$ on a open set $U$
containing $p$ as in Proposition~\ref{convex-graph} so that $\cH\cap
U$ is given by $x^{n+1}=h(x^1,\ldots,x^n)$ where $h$ is convex. We may
assume that the point $p$ has coordinates $(0,\ldots,0)$. We also
assume that $U$ is of the form $V\times (a,b)$ for $V$ an open convex
set in $=\R^n$ and that $h$ takes values in the interval $(a,b)$.
Then $h$ is locally Lipschitz and thus the Clarke differential
$\partial h(0)$ exists and is a compact convex set of linear
functionals on $\R^n$ (\cite[pp.~27--28]{Clarke:book}).  As $h$ is
convex $\partial h(0)$ is just the set of sub-differentials to $h$ at
$0$ in the sense of convex analysis (\cite[Prop.~2.2.7
p.~36]{Clarke:book}).  It follows that for $q\in \cH\cap U$ if we
write $q=(x,h(x))$ with $x\in V$ that
$$
\Nor_q(J^+(\cH))=\{\lambda(\alpha-dx^{n+1}): \lambda\ge 0, \alpha\in
\partial h(x) \}\ .
$$

There is another useful description of
$\partial h(0)$.  Let $\Omega_h$ be the set of points $x$ in $V$ where the
classical derivative $dh(x)$ exists.  As $h$ is locally Lipschitz
$\Omega_h$ has full measure in $V$.  Let $\mathcal{L}_0$ be the set 
\begin{equation}\label{L0-def}
\mathcal{L}_0:=\left\{\lim_{\ell\to\infty}dh(x_\ell):
x_\ell\in\Omega_h,\ x_\ell\to0,\ \text{and}\
\lim_{\ell\to\infty}dh(x_\ell)\ \text{exists.}\right\}
\end{equation}
of limit points of sequences $\{dh(x_\ell)\}$ of sequences
$\{x_\ell\}\subset\Omega_h$ with $x_\ell\to 0$.  Then, \cite[Thm~2.5.1
p.~63]{Clarke:book},
\begin{equation}\label{L0-hull}
\partial h(0)=\text{convex hull of}\ \mathcal{L}_0\ . 
\end{equation}

Letting, as in the introduction, $\cHdiff$ be the set of
points where $\cH$ is differentiable, if $x\in V$ and $q=(x,h(x))\in
\cH$, then $x\in \Omega_h$ if and only if $q\in \cHdiff$.  By the
theorem of~Beem and Kr{\'o}lak~\cite{BK2} this is the case if and only
if $q$ is on exactly one generator of $\cH$.  If $q=(x,h(x))\in
\cHdiff$ then let $v_q\in \cS_q$ be the unique semi-tangent to $\cH$
at $q$. Then at $q$ the tangent plane to $\cH$ can be defined either
in terms of $dh$ or in terms of $v_q$ to be the set of vectors $X\in
T_q(M)$ so that $(dh-dx^{n+1})(q)(X)=0$ or $\la v_q,X\ra=0$.  Thus
there is a positive scalar $\lambda$ so that
$(dh-dx^{n+1})(q)=\lambda\la v_q,\cdot\,\ra$.  Therefore the normal
cone at $q$ is one dimensional and
$$
\Nor_q(J^+(\cH))=\{\lambda\la v_q,\cdot\,\ra : \lambda\ge 0\}
        =\{\lambda(dh-dx^{n+1}): \lambda\ge 0\}\ .
$$
It follows from this that if $\{x_\ell\}\subset\Omega_h$ and
$q_\ell=(x_\ell,h(x_\ell))$ then $x_\ell\to 0$ if and only if
$q_\ell\to p$ and $dh(x_\ell)\to \alpha$ if and only if $v_\ell \to
v$ where $\la v,\cdot\,\ra=\alpha$.  Unraveling all this and using
\eq{L0-def} and \eq{L0-hull} gives that in order to complete the proof it
is enough to show
\begin{equation}\label{S=T}
\cS_p=\left\{\lim_{\ell\to\infty}v_{q_\ell}:q_\ell\in \cHdiff,\  q_\ell\to p,\
\text{and}\ \lim_{\ell\to \infty}v_{q_\ell}\ \text{exists.}\right\}\ .
\end{equation}
Denote the right side of this equation by $\mathcal{T}_p$.  Then,
\cite[Lemma~6.4]{Ch+}, the set of semi-tangents $\cS$ is a closed
subset of $T(M)$ and therefore $\mathcal{T}_p\subseteq \cS_p$.  If
$v\in \cS_p$ then there is a generator $c\colon [0,\infty)\to M$ with
$c(0)=p$, $c'(0)=v$ and parameterized so that it is unit speed with
respect to the auxiliary Riemannian metric $\kaux$.  For each
positive integer $\ell$, $c(1/\ell)$ is an interior point of the
generator $c$ and thus $c(1/\ell)\in \cHdiff$.  Then
$v_{c(1/\ell)}=c'(1/\ell)$ and $\lim_{\ell\to \infty}
v_{c(1/\ell)}=\lim_{\ell\to \infty}c'(1/\ell)=c'(0)=v$.  Thus $v\in
\mathcal{T}_p$ which yields $\cS_p\subseteq \mathcal{T}_p$.  This
shows~\eq{S=T} holds and completes the proof of the lemma and
therefore of Theorem~\ref{th:sing}. \qed
\end{proof}

\section{Application to distance functions and 
cut loci in Riemannian Manifolds.}

Let $(S,h)$ be a connected Riemannian manifold which we do not assume
to be complete.  Let $C\subset S$ be a closed set.  Then define the
\emph{distance function} $\rho_C\cn S\to [0,\infty)$ by
$$
\rho_C(p):=\text{infimum of lengths of smooth cuves in $S$ connecting
$p$ to $C$}\ .
$$
This will be Lipschitz with Lipschitz constant one:
$|\rho_C(p)-\rho_C(q)|\le d(p,q)$ where $d(p,q)$ is the Riemannian
distance between $p$ and $p$.

We will see that regularity properties of $\rho_C$ and the cut locus
of $C$ in $S$ are closely related to the regularity properties of
horizons, by looking at the graph of $\rho_C$ (\emph{cf.\/}
Proposition~\ref{graph=hor}).  In this setting it is natural to
consider the problem even when $(S,h)$ is not complete.  For example
when $S$ is the interior of a manifold $P$ with boundary then $\rho_C$
agrees with the distance from $C$ defined by the infimum of the length
of curves from $p$ to $C$ in $P$ so that the results apply to that
case as well.  Also in the setting of Lorentzian geometry one can use
the graphs of functions $\rho_C$ to construct examples of horizons
regardless of completeness of $(S,h)$.

Let $I\subset \R$ be an interval (which may be open, closed, or half
open).  Then a \emph{$C$-minimizing segment} on $I$ is a unit
speed geodesic $\gamma\cn I\to S$ so that
$$
\rho_C(\gamma(s))=s\quad \text{for all}\quad s\in I\ .
$$
We emphasize that we do not assume that $I$ contains $0$. The
Riemannian equivalent of the fact that horizons are
null-geodesically-ruled is contained in the following proposition:
\begin{prop}\label{min-exist}
Every $p\in S\setminus C$ is on at least one $C$-minimizing segment.
\end{prop}

\begin{proof}Let $U$ be a convex normal neighborhood of $p$,
  having closure disjoint from $C$. For $r>0$ sufficiently small the
  distance sphere $S_r(p) = \{x \in M: d(p,x) = r\}$ is contained in
  $U$, is compact, and agrees with the geodesic sphere of radius $r$
  centered at $p$.  Then, $\rho_C$ restricted to $S_r(p)$ achieves a
  minimum at some point $q$, say.  Let $c$ be the unique minimizing
  geodesic from $q$ to $p$. From the choice of $q$ on $S_r(p)$, and
  simple distance function considerations, one has for each $x$ on $c$
\begin{equation}
  \label{*}
  d(C,q) + d(q,x) = d(C,x)\;,
\end{equation}
where $d(C,x) = \rho_C(x)$.  Since $c$ is minimizing on each
segment, Equation~\eq{*} implies that $c$ is a $C$-minimizing segment.
\qed
\end{proof}

Each $C$-minimizing segment $\gamma\cn I\to S$ is contained in a
maximal (with respect to the size of the interval of definition)
$C$-minimizing segment and from now on we assume that all
$C$-minimizing segments are defined on their maximal domain.  We say
that a $C$-minimizing segment $\gamma\cn I\to S$ has a \emph{cut
  point} iff its interval of definition is of the form $[a,b]$ or
$(a,b]$ with $b<\infty$, in which case $p=\gamma(b)$ is defined to
be the cut point.  A $C$-minimizing segment can fail to have a
cut point either because its domain is unbounded, \emph{i.e.\/} of the
form $[a,\infty)$ or $(a,\infty)$, or because the domain is bounded,
say $(a,b)$ but the limit $\lim_{t\uparrow b}\gamma(t)$ does not exist
in $S$. The later condition can not arise when $S$ is complete.  When
$S$ is complete the domains of $C$-minimizing segments are all of the
form $[0,b]$ or $[0,\infty)$.

The collection of all cut points is the \emph{cut locus} of $C$ in $S$
and denoted by $\Cut_C$.  The cut locus $\Cut_C$ is a subset of
$S\setminus C$, and the definition here of $\Cut_C$ agrees with the
usual definition when $S$ is complete. For any $p\in S\setminus C$ let
$N_C(p)$ be the number of $C$-minimizing segments on which $p$ lies.
Then for all $p\in S\setminus C$ we have $N_C(p)\ge 1$ and it is known
that if $N_C(p)\ge 2$ then $p\in\Cut_C$.  For $p\in S\setminus C$ let
$$
\mathcal{M}_p:=\left\{-\gamma'(\rho_C(p)): \text{$\gamma$ is a
    $C$-minimizing segment and $\gamma(\rho_C(p))=p$}\right\}
$$
be the set of the unit vectors at $p$ which are tangent to $C$-minimizing
segments and which point toward $C$.  Then the number of vectors in
$\mathcal{M}_p$ is just $N_C(p)$.

Let $M=S\times(-\infty,0)$ and give $M$ the Lorentzian metric
$g=h-dt^2$.  Let $\cH$ be the graph of $-\rho_C$ in $M$, that is
$$
\cH:=\{(x,-\rho_C(x)): x\in S\setminus C\}\ .
$$
We leave the proof of the following to the reader.

\begin{prop}\label{graph=hor}
  The set $\cH$ is a future horizon in $M$. The null generators of
  $\cH$ are (up to reparameterization) the curves $s\mapsto
  (\gamma(s),-\rho_C(\gamma(s)))$, where $\gamma$ is $C$-minimizing
  segment of $S$ and $N(p,-\rho_C(p))=N_C(p)$.  The set $\cHend$
  coincides with the set $\{(p,-\rho_C(p)): p\in \Cut_C\}$.  \qed
\end{prop}

The theorem of Beem and Kr\'olak~\cite{BK2} that a point $p$ of a
horizon is differentiable if and only if $N(p)=1$ implies the
Riemannian result:

\begin{prop}\label{dist-diff}
With notation as above, the point $p\in S\setminus C$ is a
differentiable point of the distance function $\rho_C$ if and only if
$N_C(p)=1$.\hfill $\Box$
\end{prop}

\begin{remark}\label{diff-max}  While to best of our knowledge this result
  has not appeared in the literature on the regularity of Riemannian
  distance functions, it would be surprising if it were not known, at
  least in the case of of complete manifolds, to experts.  It can also
  be deduced from general facts about the Clarke differential in
  non-smooth analysis.  Explicitly, it follows from \cite[Thm~2.1~(4)
  p.~251]{Clarke:gen-grad}, a result about the generalized gradients
  of functions that are pointwise minimums of families of smooth
  functions with appropriate Hessian bounds, that the Clarke
  differential of $\rho_C$ at $x\in N\setminus C$ is
$$
\partial\rho_C(x)=\text{convex hull of }\{-\la u,\cdot\,\ra: u\in \mathcal{M}_x\}
$$
(When $S$ is Euclidean space this is~\cite[Lemma~4.2
pp.~1037--1038]{Fu:dist-fcn}. The extension to complete Riemannian
manifolds is not hard.)  The function $\rho_C$ is semi-convex and a
semi-convex function is differentiable at $x$ if and only its Clarke
differential at $x$ is a singleton\footnote{The general semi-convex
  case reduces to the case of convex functions.  For a convex function
  the Clarke differential is the sub-differential in the sense of
  convex analysis~\cite[Prop~2.2.7 p.~36]{Clarke:book} and a convex
  function is differentiable at point if and only if its
  sub-differential is a singleton.}.  Therefore $\rho_C$ is
differentiable at $x$ if and only if $N_C(x)=1$.  It is also possible
to carry out a proof of the Beem and Kr\'olak result along these
lines.
\end{remark}

In~\cite{Chrusciel-Galloway:horizons} horizons in Lorentzian manifolds
are constructed that are non-differentiable on a dense set.  Another
family of such examples, possessing genericity properties, is given
in~\cite{Budzynski-Kondracki-Krolak}.  When translated into Riemannian
terms the examples of~\cite{Chrusciel-Galloway:horizons} imply:

\begin{prop}\label{not-diff}
  There exists a closed Lipschitz curve $C\subset \R^2$ such that its
  distance function $\rho_C$ is non-differentiable on a dense subset
  of the set $\rho_C\le 1$ of the unbounded component of
  $\R^2\setminus C$.\hfill $\Box$
\end{prop}

In both~\cite{Chrusciel-Galloway:horizons,Budzynski-Kondracki-Krolak}
it is shown how to obtain higher dimensional examples.

Theorem~\ref{th:a} implies

\begin{prop}\label{cut:measure}
If $C$ is a closed set in an $n$-dimensional Riemannian
manifold $(S,h)$, then $\Hau^n_H(\Cut_C)=0$.\hfill $\Box$
\end{prop}
When $C$ is a smooth submanifold of $S$ (for example when $C$ is a
point) and the manifold $(S,h)$ is complete then a recent result of
Itoh and Tanaka~\cite[Thm.~B p.~22]{Itoh-Tanaka} implies that $\Cut_C$
has Hausdorff dimension at most $n-1$. (When $S$ is two dimensional
and $C$ is a point this had been done earlier by
Hebda~\cite{Hebda:cut} and Itoh~\cite{Itoh:cut}.)  However for
arbitrary closed sets the question of the Hausdorff dimension of
$\Cut_C$ is open.  In particular it is not known if there is a closed
subset $C$ of the Euclidean plane $\R^2$, with its usual metric, so
that the Hausdorff dimension of $\Cut_C$ is~$2$.
(Proposition~\ref{cut:measure} implies $\Hau^2(\Cut_C)=0$, but this
does not rule out the possibility that the Hausdorff dimension
is~$2$.)

If $M=S\times (-\infty,0)$ is given as its auxiliary Riemannian
metric $\kaux =h+dt^2$, then the set of semi-tangents to $\cH$ at
$p=(x,-\rho_C(x))$ is $\cS_p=\{ (2^{-1/2}u,2^{-1/2}\partial/\partial t):
u\in \mathcal{M}_x\}$.  Set
$$
\Cut_C[k]:=\big\{x\in S\setminus C : \dim \Span\{(u,\partial
/\partial t):
u\in \mathcal{M}_x\}\ge k\, \big\}\ .
$$
As in the introduction $\{x\in S\setminus C: N_C(p)\ge k\}\subseteq
\Cut_C[k]$ and equality holds if $1\le k\le 3$.  Note for $k\ge 2$
that $\Cut_C[k]\subset \Cut_C$, while for $k=1$ we have
$\Cut_C[1]=S\setminus C$.  The set $\{x\in N\setminus C: N_C(x)\ge
2\}=\Cut_C[2]$ is the \emph{strict cut locus} and for some special
choices of $C$, for example a point or a submanifold, its structure
has been studied by several authors ({\em cf.\/}
\cite{Hebda:cut,Itoh-Tanaka} and the references therein).  Using that
the epi-graph of a distance function has locally positive reach it
follows from results of Federer~\cite[Remark~4.15 (3)
p.~447]{Federer:measures} that each set $\Cut_C[k]$ is countable
rectifiable of dimension~$n+1-k$.  Theorem~\ref{th:sing} allows us to
refine this to $C^2$~rectifiablity.

\begin{prop}\label{cut:st}
  For $1\le k\le n+1$ the set $\Cut_C[k]$ is a $C^2$~rectifiable set
  of dimension~$n+1-k$.  Therefore $\Cut_C[k]$ has Hausdorff dimension
  $\le n+1-k$. ~ \hfill$\Box$
\end{prop}

For each $r>0$ let $\tau_r(C):=\{p\in S: \rho_C(p)=r\}$ be the tube of
radius $r$ about $C$. In general these tubes can have singularities
and need not be topological hypersurfaces in $S$.  We now look at what
the Riemannian versions of~\cite[Sec.~5]{Ch+} have to say about the
regularity of $\tau_r(C)$ and $\rho_C$.  By
Proposition~\ref{convex-graph} the function $\rho_C$ is semi-convex on
$S\setminus C$.  Therefore by Alexandrov's theorem for almost all
$p\in S\setminus C$ the function $\rho_C$ has second Alexandrov
derivatives.  This means that in local coordinates $x^1,\ldots, x^n$
centered at $p$ the function $\rho_C$ has a second order Taylor
expansion
$$
\rho_C=\rho_C(p)+d\rho_C(p)x+\frac12D^2\rho_C(p)(x,x)+o(|x|^2)
$$
where $x=(x^1,\ldots, x^n)$, $d\rho_C(p)$ is a covector at $p$ and
$D^2\rho_C(p)\cn T_p(S)\times T_p(S)\to \R$ is a symmetric bilinear
form.  Denote by $\SAl$ the set of Alexandrov points of $\rho_C$.  At
points $p\in \SAl$ the function $\rho_C$ is differentiable and
therefore the discussion above yields that $N_C(p)=1$ and that
$d\rho_C(p)=-\la u,\cdot\,\ra$ where $u\in \mathcal{M}_p$.  If $p\in
\SAl$ and $r=\rho_C(p)$ then the level set $\tau_r(C)$ has a well
defined tangent space $T_p(\tau_r(C)):=\{X\in T_pS: d\rho_C(p)X=0\}$
and a well defined second fundamental form $\ii$ at $p$ given by
$$
\ii_p(X,Y)=-D^2\rho_C(p)(X,Y)\quad \text{for}\quad X,Y\in
T_p(\tau_r(C))\ .
$$
Let $A_p\cn T_p(\tau_r(C))\to T_p(\tau_r(C))$ be the corresponding
Weingarten map defined by
$$
\la A_pX,Y\ra :=\ii_p(X,Y)\quad \text{for} \quad X,Y\in T_p(\tau_r(C))\ .
$$
That is $A_p$ is the tensor of type $(1,1)$ corresponding to the
tensor~$\ii$ of type~$(0,2)$ and $A_p$ is a self-adjoint linear map.
With the choice of signs here, when $S=\R^n$ and $C$ is the origin (so
that $\tau_r(C)$ is the sphere of radius $r$) for $p\in \tau_r(C)$ we
have $A_p=-\frac1r I$.  Let the Riemannian connection on $S$ be
denoted by $\nabla$ and let $R$ be the curvature tensor of $S$ with
signs chosen so that
$R(X,Y)Z=(\nabla_X\nabla_Y-\nabla_Y\nabla_X-\nabla_{[X,Y]})Z$.  For
$U\in T_pM$ define a linear map $R_U\cn T_pS\to T_pS$ be
$R_UX=R(X,U)U$.  The following summarizes the Riemannian versions of
Proposition~3.1, Proposition~3.5, Theorem~5.1 and Theorem~5.6
of~\cite{Ch+}.
\begin{thm}\label{thm:Alex}
  Let $\gamma\cn I\to S$ by a $C$-miminizing segment with
  $\gamma(0)\in C$.  If $\gamma(t_0)\in \SAl$ for some $t_0\in I$,
  then $\gamma(t)\in \SAl$ for all $t\in I_0:=\{t_0\}\cup (I\cap
  [t_0,\infty))^\circ$ (where $J^\circ$ is the interior of $J\subset
  \R$).  The Weingarten maps $A_{\gamma(t)}$ of the tubes
  $\tau_{\rho_C(\gamma(t))}(C)$ vary smoothly on the interval $I_0$
  and satisfy the usual Riccati equation for parallel hypersurfaces:
$$
 \nabla_{d\over dt}A_{\gamma(t)}=A_{\gamma(t)}^2+R_{\gamma'(t)}\ .
$$
 Moreover for all $r\in (0,\infty)$ the tube $\tau_r(C)$ has locally
 finite $\Hau^{n-1}$-dimensional measure and for almost all $r\in
 (0,\infty)$ there is a set $P\subset \tau_r(C)$ with $\Hau^{n-1}(P)=0$
 so that every $C$-miminizing segment $\gamma\cn I\to S$ which meets
 $\tau_r(C)\setminus P$ will have $\gamma(t)\in \SAl$ for $t\in
 I^\circ$.\hfill $\Box$
\end{thm}

In the terminology of~\cite{Ch+} ``Alexandrov points propagate to the
past along generators'' of horizons.  In the Riemannian setting
Alexandrov points of a distance function $\rho_C$ propagate away from
$C$ along $C$-miminizing segments.  The last sentence of
Theorem~\ref{thm:Alex} implies loosely that almost every $C$-miminizing
segment is an \emph{Alexandrov segment} in the sense that all of its
points other than its endpoints are in $\SAl$.


\begin{thebibliography}{10}

\bibitem{Alberti:sing}
G.~Alberti.
\newblock On the structure of singular sets of convex functions.
\newblock {\em Calc. Var. Partial Differential Equations}, 2(1):17--27, 1994.

\bibitem{BK2}
J.~K. Beem and A.~Kr{\'o}lak.
\newblock Cauchy horizon end points and differentiability.
\newblock {\em J. Math. Phys.}, 39(11):6001--6010, 1998.

\bibitem{Budzynski-Kondracki-Krolak}
R.~J. Budzy{\'n}ski, W.~Kondracki, and A.~Kr{\'o}lak.
\newblock On the differentiability of {C}auchy horizons.
\newblock {\em J. Math. Phys.}, 40(10):5138--5142, 1999.

\bibitem{Ch+} P.~T. Chru{\'s}ciel, E.~Delay, G.~J. Galloway, and
  R.~Howard.  \newblock Regularity of horizons and the area theorem.
  \newblock {\em To appear in \emph{Annales H.~Poincar\'e}},
  gr-qc/0001003, 2000.

\bibitem{Chrusciel-Galloway:horizons}
P.~T. Chru{\'s}ciel and G.~J. Galloway.
\newblock Horizons non-differentiable on a dense set.
\newblock {\em Comm. Math. Phys.}, 193(2):449--470, 1998.

\bibitem{Clarke:gen-grad}
F.~H. Clarke.
\newblock Generalized gradients and applications.
\newblock {\em Trans. Amer. Math. Soc.}, 205:247--262, 1975.

\bibitem{Clarke:book}
F.~H. Clarke.
\newblock {\em Optimization and nonsmooth analysis}.
\newblock Society for Industrial and Applied Mathematics (SIAM), Philadelphia,
  PA, second edition, 1990.

\bibitem{Federer:measures}
H.~Federer.
\newblock Curvature measures.
\newblock {\em Trans. Amer. Math. Soc.}, 93:418--491, 1959.

\bibitem{Federer:book}
H.~Federer.
\newblock {\em Geometric measure theory}.
\newblock Springer-Verlag New York Inc., New York, 1969.
\newblock Die Grundlehren der mathematischen Wissenschaften, Vol.~153.

\bibitem{Fu:dist-fcn}
J.~H.~G. Fu.
\newblock Tubular neighborhoods in {E}uclidean spaces.
\newblock {\em Duke Math. J.}, 52(4):1025--1046, 1985.

\bibitem{Hebda:cut}
J.~J. Hebda.
\newblock Metric structure of cut loci in surfaces and {A}mbrose's problem.
\newblock {\em J. Differential Geom.}, 40(3):621--642, 1994.

\bibitem{HusaWinicour}
S.~Husa and J.~Winicour.
\newblock Asymmetric merger of black holes.
\newblock {\em Phys. Rev. D (3)}, 60(8):084019, 13, 1999.
\newblock \emph{gr-qc/9905039}.

\bibitem{Itoh:cut}
J.~Itoh.
\newblock The length of a cut locus on a surface and {A}mbrose's problem.
\newblock {\em J. Differential Geom.}, 43(3):642--651, 1996.

\bibitem{Itoh-Tanaka}
J.~Itoh and M.~Tanaka.
\newblock The {L}ipschitz continuity of the distance function to the cut locus.
\newblock {\em Trans. Amer. Math. Soc.}, 353(1):21--40, 2000.

\end{thebibliography}
\end{document}